\documentclass[a4paper,showpacs,prb,twocolumn]{revtex4}
\usepackage{amsfonts}
\usepackage{amsmath}
\usepackage{amssymb}
\usepackage{graphicx}
\usepackage{verbatim}

\begin{document}

\title{Capacitance of graphene nanoribbons}
\author{A. A. Shylau$^{1}$, J. W. K{\l }os$^{1,2}$, and I. V. Zozoulenko$%
^{1} $}
\affiliation{$^{1}$Solid State Electronics, ITN, Link\"{o}ping University, 601 74, Norrk%
\"{o}ping, Sweden\\
$^{2}$Surface Physics Division, Faculty of Physics, Adam Mickiewicz
University, Umultowska 85, 61-614 Poznan, Poland}

\begin{abstract}
We present an analytical theory for the gate electrostatics and the
classical and quantum capacitance of the graphene nanoribbons (GNRs)
and compare it with the exact self-consistent numerical calculations
based on the tight-binding $p$-orbital Hamiltonian within the
Hartree approximation. We demonstrate that the analytical theory is
in a good qualitative (and in some aspects quantitative) agreement
with the exact calculations. There are however some important
discrepancies. In order to understand the origin of these
discrepancies we investigate the self-consistent electronic
structure and charge density distribution in the nanoribbons and
relate the above discrepancy to the inability of the simple
electrostatic model to capture the classical gate electrostatics of
the GNRs. In turn, the failure of the classical electrostatics is
traced to the quantum mechanical effects leading to the significant
modification of the self-consistent charge distribution in
comparison to the non-interacting electron description. The role of
electron-electron interaction in the electronic structure and the
capacitance of the GNRs is discussed. Our exact numerical
calculations show that the density distribution and the potential
profile in the GNRs are qualitatively different from those in
conventional split-gate quantum wires; at the same time, the
electron distribution and the potential profile in the GNRs show
qualitatively similar features to those in the cleaved-edge
overgrown quantum wires. Finally, we discuss an experimental
extraction of the quantum capacitance from experimental data.
\end{abstract}

\date{\today }
\maketitle

\section{Introduction}

Graphene, a two-dimensional (2D) honeycomb structure of carbon atoms, has
attracted a lot of interest since its isolation in 2004\cite{Novoselov}. It
demonstrates unique properties which originate from the Dirac-type spectrum
of low-energy quasiparticles. Nowadays, graphene is considered to be a
viable alternative to Si for the channel of field-effect transistors (FETs)%
\cite{FET}. One of the main characteristics of such devices is a capacitance
formed between the channel and the gate. The capacitance is important for
understanding fundamental electronic properties of the material such as the
density of states (DOS) as well as device performance including the $I-V$
characteristics and the device operation frequency.

In a classical regime, the capacitance describes the capability of an object
to store electrical charges and is completely determined by the object's
geometry and a dielectric constant of the medium. If the object's size
shrinks to a nanometer scale, quantum effects have to be taken in account.
One of manifestations of these effects is a finite DOS which originates from
the Pauli exclusion principle. Low-dimensional systems, having a small DOS,
are not able to accumulate enough charge to completely screen the external
field. In order to describe the effect of the electric field penetration
through a two-dimensional electronic gas (2DEG) Luryi introduced a concept
of a quantum capacitance\cite{Luryi}.

Recently, the quantum capacitance of a bulk graphene layer deposited on a
gated SiO$_{2}$ insulated surface has been investigated by means of scanning
probe microscopy\cite{Giannazzo}. To the best of our knowledge, no studies
of the gate capacitance of the graphene nanoribbons (GNRs) have been
reported yet. However, such studies are already technologically feasible.
Indeed, during last years the great progress has been achieved in
fabrication and patterning of the GNRs\cite{Chen,Han,Li,Molitor} as well as
in controlling the morphology, geometry and stability of the graphene edges%
\cite{Jia,Girit}. On the other hand, the quantum and classical
capacitance of related structures, - carbone nanotubes, has been
measured and analyzed by a number of groups during the last
years\cite{Ilani,Liang,Dai,Pomorski}. The later studies have
revealed a number of interesting properties of the system at hand
including the structure of the DOS and signatures of the electron
interaction and correlation.

In order to provide physical insight into the gate electrostatics
and capacitance of the GNRs, it is important to develop intuitive
analytical models capturing the essential physics of the device at
hand. Such models are also imperative in experimental measurements
because the quantum capacitance is not directly accessible in the
experiments and can only be indirectly extracted from the measured
total capacitance. In the present paper we develop a basic
analytical theory for the gate electrostatics and the classical and
quantum capacitance of the GNRs. We complement this analytical
theory by exact self-consistent numerical calculations based on the
tight-binding $p$-orbital Hamiltonian within the Hartree
approximation. We demonstrate that the analytical theory is in a
good qualitative (and in some aspects quantitative) agreement with
the exact calculations. There are however some important
discrepancies. In order to understand the origin of these
discrepancies we investigate the self-consistent electronic
structure and charge density distribution in the nanoribbons and
relate the above discrepancy to the inability of the simple
electrostatic model to capture the classical gate electrostatics of
the GNRs. In turn, the failure of the classical electrostatics is
traced to the quantum mechanical effects leading to the significant
modification of the self-consistent charge distribution in
comparison to the non-interacting electron description.

It should be noted that particular aspects of the self-consistent gate
electrostatics and the electron structure of the GNRs\cite{Fernandez-Rossier}
and the numerical\cite{Guo} and analytical\cite{Fang} studies of quantum
capacitance of the GNRs have been reported in the literature. In particular,
the quantum capacitance of the GNRs as a function of the Fermi
energy,\thinspace $C_{Q}=C_{Q}(E_{F}),$ has been studied in Ref. [\cite{Fang}%
]. Experimentally, however, the dependence of $C_{Q}$ on the Fermi energy is
not accessible, and we stress that the focus of our analytical and numerical
analysis is the gate capacitance, $C=C(V_{g}),$ - the characteristic that is
measured experimentally ($V_{g}$ being the gate voltage).

The paper is organized as follows. In Sec. II we formulate the basics of our
model of the gated GNRs. The analytical treatment of the gate electrostatics
and the quantum and classical gate capacitance of GNRs is given in Sec. III.
The results of the self-consistent numerical calculations and a comparison
between the analytical and numerical calculations are presented and
discussed in Sec. IV. Section V summarizes the main conclusions.

\section{Model}

In experiments graphene samples are separated from the gate by a relatively
thick insulating substrate (of a typical width of at least $300$ nm) in
order to enable visual identification of the graphene sheet. This simplest
experimental setup (with a single back gate) is not particularly suitable
for measurements of the quantum capacitance because in this case (as we will
demonstrate below) the total capacitance is completely dominated by the
classical contribution and can hardly be extracted from the measured total
capacitance. In order to distinguish the quantum contribution the gate
should be placed much closer to the ribbon such that the classical
capacitance $C_{C}$ becomes comparable to the quantum one. In our study we
therefore consider an embedded top-gate geometry shown in Fig. \ref%
{fig:geometry} where a graphene ribbon of the width $w$ is placed on a thick
dielectric layer and covered by the second much thinner layer of the width $%
d $ separating it from the top gate with the applied gate voltage $V_{g}$.
\begin{figure}[tbh]
\includegraphics[keepaspectratio,width=\columnwidth]{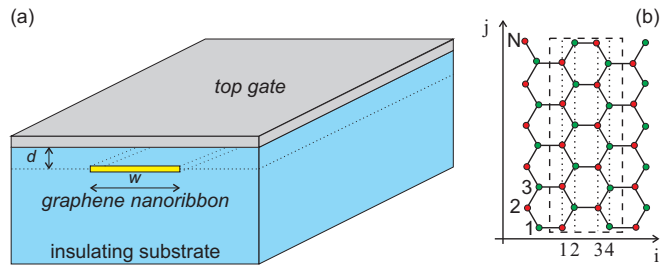}
\caption{(Color online) (a) A schematic diagram illustrating a top-gate
geometry where an infinitely long graphene nanoribbon of the width $w$ is
embedded in a gate insulator with the relative dielectric constant $\protect%
\epsilon _{r}$; $d$ is the distance between the ribbon and the top
gate. It is assumed that the graphene nanoribbon is connected to the
source and drain reservoirs supplying electrons to the ribbon. (b)
An armchair graphene ribbon of the width $N$. All the results
presented in this paper correspond to $N=98$ ($w=12$ nm.)}
\label{fig:geometry}
\end{figure}
This top-gate geometry was used by Ilani \textit{et al.}\cite{Ilani} and
Natori \textit{et al.}\cite{Natori} for measurements of the quantum
capacitance of carbone nanotubes. We also assume that the GNR is connected
to the source and drain electrods playing a role of ideal reservoirs
supplying electrons to the ribbon. The experimental setup might also include
the back gate which can, independently of the top gate, adjust a position of
the Dirac point in the graphene nanoribbon and hence change its electron
density. However we assume that the back gate is situated much further apart
from the ribbon in comparison to the top gate, and therefore for the sake of
simplicity we in our model disregard a possible capacitative coupling
between the ribbon and the back gate.

In this paper we considered two representative structures, (I) HfO$_{2}$
insulating layer with $d=30$ nm and $\varepsilon _{r}=47$ \cite%
{Park,Ozyilmaz,Fernandez-Rossier}, and (II) SiO$_{2}$ insulating layer with $%
d=300$ nm and $\varepsilon _{r}=3.9$. For the first structure
$C_{C}$ is comparable to $C_{Q},$ whereas for the second one
$C_{C}\ll C_{Q}$. We limit our calculations to the case of the
armchair GNRs, whereas we expect that the main results and
conclusions presented in this paper can be extended to the case of
the zigzag GNRs. (Note that we do not focus on any specific
peculiarities of the DOS near the Dirac point like surface states in
the case of zigzag ribbons). The spin effects and the effect of
disorder are outside the scope of our paper and are deferred to
future studies. All
results correspond to the metallic armchair GNRs with the width $w=12$ nm ($%
N=98$). We also made calculations for a semiconductor armchair GNR as well
as for wider ribbons ($w=$50 nm) and all the results show the same features.

The system presented on Fig. \ref{fig:geometry} is described by the standard
$p$-orbital tight-binding Hamiltonian\cite{RMP,Dresselhaus}
\begin{equation}
H=\sum_{\mathbf{r}}V_{H}(\mathbf{r})a_{\mathbf{r}}^{+}a_{\mathbf{r}}-\sum_{%
\mathbf{r},\Delta }t_{\mathbf{r},\mathbf{r}+\Delta }a_{\mathbf{r}}^{+}a_{%
\mathbf{r}+\Delta },  \label{Htb}
\end{equation}%
where $t_{\mathbf{r},\mathbf{r}+\Delta }=2.5$ eV is a nearest-neighbor
hopping integral; $V_{H}(\mathbf{r})$ is a Hartree potential at the site
\textbf{r} which results from the Coulomb interaction between extra charges $%
q(\mathbf{r})$ in the system (including the mirror charges)\cite%
{Fernandez-Rossier,Ihnatsenka,cleaved_edge},
\begin{equation}
V_{H}(\mathbf{r})=-\frac{e^{2}}{4\pi \varepsilon _{0}\varepsilon _{r}}\sum_{%
\mathbf{r}^{^{\prime }}\neq \mathbf{r}}q(\mathbf{r}^{^{\prime }})\left(
\frac{1}{|\mathbf{r}-\mathbf{r}^{^{\prime }}|}-\frac{1}{\sqrt{|\mathbf{r}-%
\mathbf{r}^{^{\prime }}|^{2}+4d^{2}}}\right) ,  \label{Vr}
\end{equation}%
The summation in Eq. (\ref{Vr}) can be split into two parts corresponding to
the $A$ and $B$ sublattices of graphene (see Fig. \ref{fig:geometry} (b)).
Changing summation to integration in the $i$-direction we obtain,
\begin{eqnarray}
V_{H}(\mathbf{r}_{ij}) &=&-\frac{e^{2}}{4\pi \varepsilon _{0}\varepsilon _{r}%
}\sum_{j^{\prime }=1\,(j^{\prime }\neq j)}^{N}\frac{q_{A}(j^{\prime
})+q_{B}(j^{\prime })}{2}  \label{Vh} \\
&&\times \ln \frac{(\mathbf{r}_{ij}-\mathbf{r}_{ij^{\prime }})^{2}}{(\mathbf{%
r}_{ij}-\mathbf{r}_{ij^{\prime }})^{2}+4d^{2}},  \notag
\end{eqnarray}%
where $q_{A(B)}(j^{\prime })$ is the charge on the carbon atom which is
located on the $j^{\prime }$ line and corresponds to the $A(B)$ sublattice.

We solve Eq. (\ref{Htb}) numerically to find the Green's function
using the technique described by Xu \emph{et al.} \cite{Xu}. This
technique greatly facilitates computation speed since it does not
require self-consistent calculation of the surface Green's function.
The Green's function in the real-space representation,
$G(\mathbf{r},\mathbf{r})$, provides an information about the local
density of states (LDOS) at site \textbf{r},
\begin{equation}
\rho (\mathbf{r},E)=-\frac{2}{\pi S}\Im \lbrack
(G(\mathbf{r},\mathbf{r}))], \label{LDOS}
\end{equation}%
where factor 2 indicates a spin degeneracy and $S$ is the area corresponding
to one carbon atom. The LDOS can be used to calculate the local electron
density at the site \textbf{r},
\begin{equation}
n(\mathbf{r},E_{F})=\int_{eV_{C}}^{E_{F}}dE\rho (\mathbf{r}%
,E)f_{FD}(E-E_{F}),  \label{n_sc}
\end{equation}%
where $E_{F}=eV_{g}$ is Fermi energy and $f_{FD}$ is the Fermi-Dirac
distribution function. (All the calculations reporeted in this paper
correspond to the temperature $T=0$ K). The position of the charge
neutrality point $eV_{C}$ at a given gate voltage $V_{g}$ is determined
numerically from the calculated dispersion relation. For example, for the
armchair GNRs, the position of the charge neutrality point $eV_{C}$
corresponds to the energy which gives the minimum number of propagating
states with the smallest absolute value of the wave vector. Note, that in
order to achieve a fast convergence, the itegration in Eq. (\ref{n_sc}) is
performed in a complex plane, since on the real $E$-axis $\rho (\mathbf{r}%
,E) $ is a rapidly varying function of the energy (see Refs. \cite%
{Ihnatsenka,cleaved_edge} for details) .

Since the Hartree potential $V_{H}$ (\ref{Vh}) depends on the electron
density $n(\mathbf{r})$ which is a solution of the Schr\"{o}dinger equation
with the Hamiltonian (\ref{Htb}), these equations need to be solved
iteratively. The iteration process is executed until the convergence
criterion is met, $\left\vert \frac{V_{out}^{m}-V_{in}^{m}}{%
V_{out}^{m}+V_{in}^{m}}\right\vert <10^{-5}$, where $V_{in}^{m}$ and $%
V_{out}^{m}$ are the input and output average values of the Hartree
potential on the $m$-th iteration. In order to accelerate convergence we
used the Broyden's second method\cite{Broyden}, which allows us to reduce
the number of iterations to $\sim {8-10}$ in comparison to $\sim {40-50}$
iterations needed with the "simple mixing" method.

Having calculated the electron density and the position of the Dirac
point numerically, we are in position to find the total, quantum and
classical capacitances as a function of the gate voltage, see Sec.
IV for details. The analytical approach to the quantum and classical
capacitance of the GNRs is described in the next section.

\section{Classical and quantum capacitance of graphene nanoribbons: an
analytical model}

\begin{figure}[t]
\includegraphics[keepaspectratio,width=0.5\columnwidth]{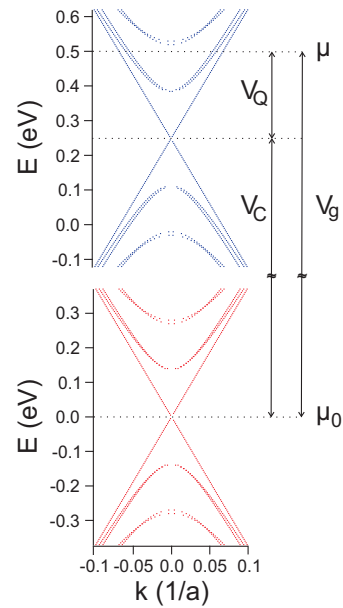}
\caption{(Color online) A diagram illustrating the change in the band
structure of the graphene nanoribbon and a shift of the chemical potential $%
\protect\mu $ under the application of the gate voltage $V_g$ (see text for
details). The displayed diagram corresponds to the HfO$_2$ structure with $%
d=30$ nm and $V_g=0.5$ V.}
\label{fig:band_evolution}
\end{figure}

The electronic structure of planar gated graphene sheets was studied by Fern%
\'{a}ndez-Rossier \textit{et al}. \cite{Fernandez-Rossier}. In this section
we follow the approach outlined by Fern\'{a}ndez-Rossier \textit{et al}.
\cite{Fernandez-Rossier} and provide an analytical description of the
one-dimensional charge density and quantum and classical capacitance of the
graphene nanoribbons.

Application of the gate voltage $V_{g}$ to a metallic gate induces extra
carriers with the density $n$ to the nanoribbon as well as extra carriers of
the opposite polarity to the gate itself. (Note that the gate and the ribbon
represent together a charge neutral system). The application of the gate
voltage shifts the chemical potential $\mu $ of the ribbon from the charge
neutrality point $\mu _{0}$ \cite{Fernandez-Rossier},%
\begin{equation}
eV_{g}=\mu -\mu _{0},  \label{mu1}
\end{equation}%
It is convenient to represent this shift as a sum of two terms%
\begin{equation}
\mu -\mu _{0}=eV_{C}+eV_{Q},  \label{mu2}
\end{equation}%
where $eV_{C}$ describes the position of the charge neutrality point at the
applied voltage $V_{g},$ and $eV_{Q}$ describes the change in the chemical
potential due to the filling of the quantum mechanical energy bands, see
Fig. \ref{fig:band_evolution} for illustration. For a classical conductor
the density of states is infinite and thus $V_{Q}=0.$ This provides a
natural interpretation of $V_{C}$ as a classical electrostatic potential,
whereas the potential $V_{Q}$ has a quantum mechanical origin reflecting the
structure of the quantum mechanical density of states. Using a relation
\begin{equation}
V_{g}=V_{Q}+V_{C}  \label{V_g}
\end{equation}%
which follows from Eqs. (\ref{mu1}) and (\ref{mu2}) and using a definition
of a capacitance $C=\frac{e\partial n}{\partial V},$ we obtain\cite%
{Fernandez-Rossier}%
\begin{equation}
C_{tot}^{-1}=C_{C}^{-1}+C_{Q}^{-1},  \label{def_C}
\end{equation}%
where the total capacitance $C_{tot}=\frac{e\partial n}{\partial V_{g}},$
and the classical and the quantum mechanical capacitances are respectively $%
C_{C}=\frac{e\partial n}{\partial V_{C}}$ and $C_{Q}=\frac{e\partial n}{%
\partial V_{Q}}.$

In order to find the total capacitance for a given gate voltage we
have to calculate the electron density $n.$ The later at the zero
temperature is given by $n=\int_{E_{b}}^{\mu }\rho
(E)dE-\int_{E_{b}}^{\mu _{0}}\rho _{0}(E)dE,$ where $\rho _{0}(E)$
is the density of states for a charge neutral ribbon ($\mu =\mu
_{0}),$ $\rho (E)$ is the DOS at $\mu =\mu
_{0}+eV_{g},$ and the integration starts from the bottom of the energy band $%
E_{b}$. (In the following we will refer to the case $V_{g}=0$ as to
an uncharged ribbon, and to the case of $V_{g}\neq 0$ as a charged
ribbon). Neglecting changes in the DOS of the charged ribbon under
the applied gate voltage in comparison to the uncharged one,
\begin{equation}
\rho (E)=\rho _{0}(E-eV_{C}),  \label{mu3}
\end{equation}%
the density of the extra carriers reads\cite{Fernandez-Rossier}
\begin{equation}
n=\int_{\mu _{0}}^{\mu _{0}+eV_{Q}}\rho _{0}(E)dE.  \label{n}
\end{equation}%
(In the next Section we will demonstrate that this approximation holds
extremely well despite of some modifications of the band structure for
higher gate voltages).

The DOS of graphene armchair nanoribbons can be written in the form (see
Appendix I)%
\begin{equation}
\rho _{0}(E)=\frac{4}{\pi \sqrt{3}ta}\sum_{n}\frac{\left\vert E\right\vert }{%
\sqrt{E^{2}-E_{n}^{2}}}\theta (\left\vert E\right\vert -\left\vert
E_{n}\right\vert ),\;  \label{ro_0}
\end{equation}%
where $n=0,\pm 1,\pm 2,\ldots ,$ $t\approx 2.5$ eV is the first-neighbor
hopping integral, $a=0.246$ nm is the graphene lattice constant (note that $%
\frac{\sqrt{3}}{2}ta=v_{F}\hbar $ with $v_{F}$ being the Fermi velocity),
and $E_{n}$ are the subband threshold energies whose analytical expressions
are provided by Onipko\cite{Onipko} (see Appendix I for the explicit
expressions for $E_{n}$). Substituting this expression of the DOS into Eq. (%
\ref{n}), we obtain for the one-dimensional (1D) electron density of
the nanoribbon,
\begin{equation}
n(V_{Q})=\frac{4}{\pi \sqrt{3}ta}\sum_{n}\sqrt{(eV_{Q})^{2}-E_{n}^{2}\,}%
\theta (\left\vert eV_{Q}\right\vert -\left\vert E_{n}\right\vert ),\;
\label{n2}
\end{equation}%
where $n=0,\pm 1,\pm 2,\ldots .$ Here and hereafter without loss of
generality we set $\mu _{0}=0.$ Using the definition of the quantum
capacitance we get \cite{Fang},%
\begin{equation}
C_{Q}=\frac{e\partial n}{\partial V_{Q}}=e^{2}\rho _{0}(eV_{Q}).  \label{C_Q}
\end{equation}%
In order to calculate $C_{Q}$ from the above equation we have to know the
position of the chemical potential with respect to the charge neutrality
point of the charged ribbon, $eV_{Q}$. This can be done from Eq. (\ref{V_g}%
), where the classical electrostatic potential can be easily calculated
using the standard method of images,
\begin{equation}
V_{C}=\frac{\sigma }{\pi \varepsilon }\left[ 2d\arctan \frac{w}{4d}+\frac{a}{%
4}\ln \left\{ 1+\left( \frac{4d}{w}\right) ^{2}\right\} \right] ,
\label{V_C}
\end{equation}%
(the definitions of $w$ and $d$ are given in Fig.
\ref{fig:geometry}). In the derivation of this expression we assumed
that the surface charge density of the graphene nanoribbon $\sigma
=n/w$ is constant (as expected for a classical capacitor). We will
discuss the validity of this assumption in the next Section. The
classical capacitance (per unit length) of the graphene
nanoribbon follows from Eq. (\ref{V_C}),%
\begin{equation}
C_{C}=\frac{e\partial n}{\partial V_{C}}=\pi \varepsilon w\left[ 2d\arctan
\frac{w}{4d}+\frac{a}{4}\ln \left\{ 1+\left( \frac{4d}{w}\right)
^{2}\right\} \right] ^{-1}.  \label{C_C}
\end{equation}%
Note that in the limit of a narrow ribbon, $w\ll d,$the above expression
simplifies to $C_{C}=2\pi \varepsilon /\ln \frac{4d}{w}.$

To summarize, Eqs. (\ref{V_g}), (\ref{def_C}),
(\ref{n2})-(\ref{C_C}) provide the analytical expressions for the 1D
electron density and the total, quantum and classical capacitances
of the graphene nanoribbons. In order to express the density and the
capacitances as a function of the gate voltage $V_{g}$ (rather than
$V_{Q}$ which is not accessible experimentally) we first choose some
value of $V_{Q}$ and calculate $n$ and $C_{Q}$ from
Eqs. (\ref{n2}), (\ref{C_Q}). We then use the calculated values in Eqs. (\ref%
{V_C}),(\ref{C_C}) to find corresponding $V_{C}$ and $C_{C}$. Finally,
relating $V_{Q}$ to $V_{g}$ via Eq. (\ref{V_g}), we express the density and
the capacitance as a function of the gate voltage $V_{g}.$

\section{Results and discussion}

\begin{figure}[t]
\includegraphics[keepaspectratio,width=\columnwidth]{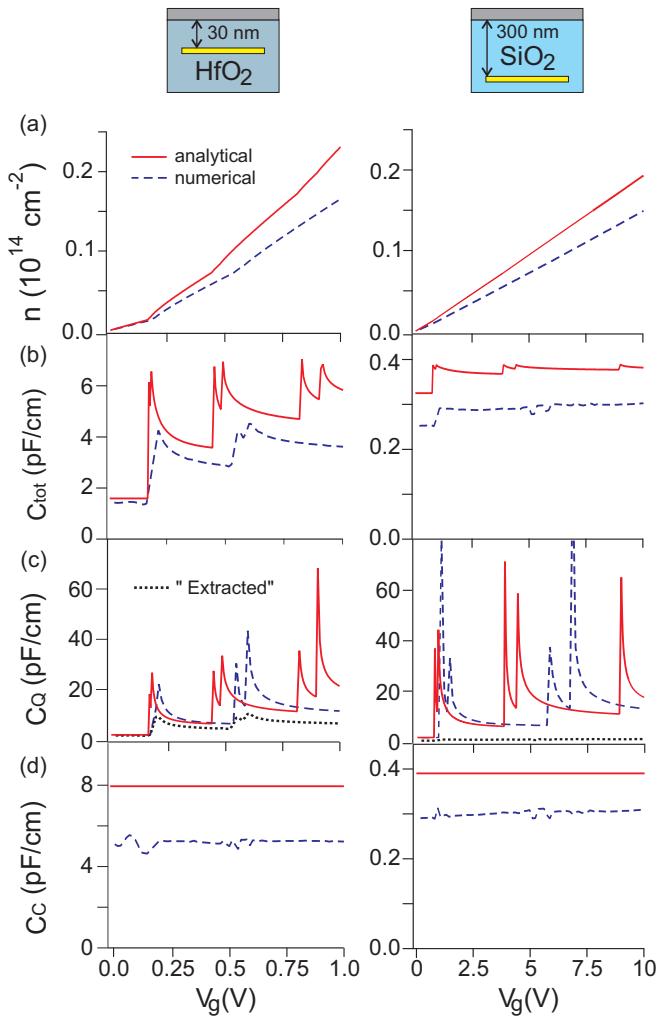}
\caption{(Color online) The analytical and numerical dependencies on the
applied gate voltage of (a) the electron density $n$, (b) the total
capacitance $C_{tot}$, (c) the quantum capacitance $C_{Q}$, (d) the
classical capacitance $C_{C}$. (c) also shows the ``extracted" quantum
capacitance (see text for details). Left and right panels corresponds
respectively to HfO$_{2}$ and SiO$_{2}$ structures.}
\label{fig:capacitance}
\end{figure}

Figure \ref{fig:capacitance} shows the analytical and numerical densities
and capacitances for two representative nanoribbon structures introduced in
Sec. II. The left column corresponds to the 30 nm HfO$_{2}$ dielectric
structure with $\varepsilon _{r}=47,$ and the right column corresponds to a
conventional 300 nm SiO$_{2}$ structure with $\varepsilon _{r}=3.9.$ The
analytical results are based on the expressions given by Eq. (\ref{V_g}), (%
\ref{def_C}), (\ref{n2})-(\ref{C_C}) and show the electron density $n,$ the
total, quantum and classical capacitances $C_{tot},C_{Q},C_{C}$ (Figs. \ref%
{fig:capacitance} (a), (b), (c), (d) respectively).

The numerical results are based on the self-consistent solution of Eqs. (\ref%
{Htb})-(\ref{n_sc}) as described in Sec. II. We first calculate the electron
density as a function of the gate voltage and then differentiate it
numerically in order to compute the total capacitance, $C_{tot}=\frac{%
e\partial n}{\partial V_{g}},$ see Figs. \ref{fig:capacitance} (a) and (b)
respectively. Figure \ref{fig:capacitance} (c) shows the quantum capacitance
$C_{Q}$ which is calculated from the DOS on the basis of Eq. (\ref{C_Q})
(Note that $\rho _{0}(E)=\frac{1}{N_{cell}}$ $\sum_{\mathbf{r}}\rho (\mathbf{%
r},E)$, where $\rho (\mathbf{r},E)$ is the LDOS given by Eq.
(\ref{LDOS}), and summation is performed over one unit cell
containing $N_{cell}$ sites). Having calculated $C_{tot}$ and
$C_{Q}$ we compute the classical capacitance
from Eq. (\ref{def_C}) as $C_{C}^{-1}=C_{tot}^{-1}-C_{Q}^{-1},$ see Fig. \ref%
{fig:capacitance} (d).

The total capacitance $C_{tot}$ for both structures shows characteristic
features that can be traced to the corresponding features in the quantum
capacitance $C_{Q}$, (cf. Figs. \ref{fig:capacitance} (b) and (c)). Because
the quantum capacitance is proportional to the DOS (see Eq. (\ref{C_Q})) the
peaks in $C_{Q}$ signal consecutive population of electron subbands as the
gate voltage increases. Note that these features in $C_{tot}$ are much less
pronounced for the case of a conventional SiO$_{2}$ structure because its
classical capacitance is much smaller than the quantum one ($C_{Q}/C_{C}\sim
40),$ whereas for the HfO$_{2}$ structure this ratio is only $\sim 2$ (note
that the quantum and classical capacitances are added in series, Eq. (\ref%
{def_C})).

The comparison of the analytical and numerical calculations demonstrates
that the analytical theory qualitatively reproduces the exact results very
well. There is however some quantitative discrepancy in the values of both
quantum and classical capacitances. In particular, the analytical classical
capacitance differs by 20-35\% from its exact numerical value, see Fig. \ref%
{fig:capacitance} (d). For the case of the SiO$_{2}$ structure the numerical
$C_{C}$ shows a slow increase as $V_{g}$ increases. This is in apparent
contrast with the behaviour of its analytical counterpart which is
independent on the applied voltage. As far as the quantum capacitance is
concerned, a visual inspection of Fig. \ref{fig:capacitance} (c) indicates
that the analytical and numerical $C_{Q}$ would coincide if one stretches
the scale of $V_{g}$ for the analytical capacitance (alternatively contracts
the scale of $V_{g}$ for the numerical capacitance).

\begin{figure}[t]
\includegraphics[keepaspectratio, width = \columnwidth]{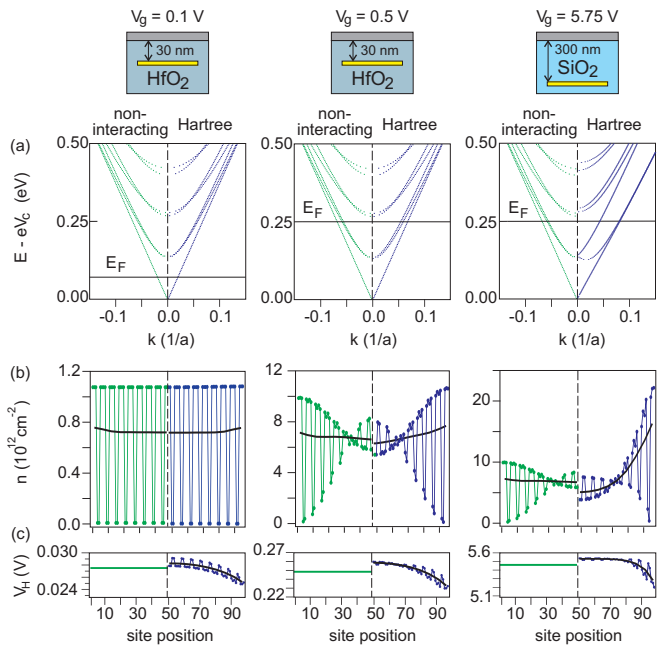}
\caption{(Color online) A comparison between the non-interacting
case (left parts of the panels) and the self-consistent Hartree
model (right parts) for the HfO$_2$ structure with $V_g=0.1$ V
(first column), HfO$_2$ structure with $V_g=0.5$ V (second column),
and SiO$_2$ structure with $V_g=5.75$ V (third column). (a) The band
structure of the nanoribbons; thin solid lines indicate the
positions of the Fermi energy. (b) The electron density distribution
across the nanoribbon. (c) The potential profiles in the Hartree
approximation (right parts); the numerically calculated value of
$V_C$ (left parts). In (b) and (c) the black lines indicate the
values of the electron density and the Hartree potential averaged
over several neighboring sites.} \label{fig:density}
\end{figure}

In order to understand the differences between the analytical and the exact
results let us critically inspect the assumptions that have been made in the
derivation of the analytical expressions in the previous section. Let us
start with the classical capacitance $C_{C}$ given by Eq. (\ref{C_C}). In
its derivation we assumed that the induced charge density is homogeneous and
the potential of the ribbon is constant as expected for a classical
conductor. Figures (\ref{fig:density}) (b),(c) show respectively the
electron density distributions and the Hartree potential for the various
nanoribbon structures for different values of $V_{Q}$ and thus for different
densities $n$ (Note that the amount of the induced charge density $n$ is
completely determined by the value of $V_{Q}$ defining the position of the
Fermi energy with respect to the charge neutrality point, Eqs. (\ref{n}), (%
\ref{n2}). For a reference purpose, the values of $V_{Q}$ are
indicated at the corresponding dispersion relations shown in Fig.
\ref{fig:density} (a) ). In order to outline the role of the
electron-electron interaction we show both the self-consistent
Hartree and the noninteracting one-electron calculations
(respectively right and left parts of the panels in Fig.
\ref{fig:density}).

The electron density distribution in the GNRs shows the pronounced
oscillation between neighboring sites. We therefore also show the electron
density averaged over several neighboring sites, Fig. \ref{fig:density} (b).
For a small gate voltage (HfO$_{2}$ structure, $V_{g}=0.1$V) when only the
first subband is filled the averaged electron density distribution is almost
uniform and there is practically no difference between the self-consistent
and the one-electron approaches. By increasing the gate voltage to $%
V_{g}=0.5V$ more carriers are induced and the electron density
increases near the edges of the structure due to the electrostatic
repulsion. It is important to stress that it is not only the
concentration of the induced charge but primarily the applied gate
voltage that determine the charge density distribution. Figure
\ref{fig:density} (b) shows the electron density distribution and
the potential profile for the SiO$_{2}$ structure for the gate
voltage $V_{g}=5.75$V. This gate voltage is chosen such that the
value of $V_{Q}=0.25$ eV is the same as for the HfO$_{2}$ structure
with $V_{g}=0.5$V (see Fig. \ref{fig:density} (a)), i.e. the induced
charge concentrations are similar. However, the density distribution
profile for the SiO$_{2}$ structure is strikingly different showing
a strong redistribution of the charges toward the edges when the
applied voltage is increased. The larger the applied voltage, the
stronger the redistribution
of the electron density. This explains the observation that the numerical $%
C_{C}$ gradually changes when the gate voltage is increased (see Fig. \ref%
{fig:capacitance} (d), right panel). We therefore conclude that the
assumptions appropriate for a classical capacitor (the charge density is
homogeneous and the potential of the ribbon $V_{C}$ is constant) are
violated for the graphene nanoribbons which leads to the difference between
the analytical theory and the exact numerical calculations.

Note that the macroscopic charge accumulation along the boundaries of the
graphene strip was discussed by Silvestrov and Efetov\cite{Efetov}. Their
semiclassical approach and the exact numerical calculations presented here
demonstrate that the density distribution and the potential profile in the
GNRs are qualitatively different from those in conventional split-gate
quantum wires with a smooth electrostatic confinement where the potential is
rather flat and the electron density is constant throughout the wire\cite%
{Ihnatsenka}. At the same time, the electron distribution and the potential
profile in the GNR are very similar to those in the cleaved-edge overgrown
quantum wires (CEOQW). Indeed, the potential profile in the CEOQWs also
exhibits triangular-shaped quantum wells in the vicinity of the wire
boundaries and the electron density is also strongly enhanced close to the
edges\cite{cleaved_edge}. This similarity simply reflects the fact that both
the CEOQWs and the GNRs correspond to the case of the hard-wall confinement
at the edges of both structures.
\begin{figure}[t]
\includegraphics[keepaspectratio,width=\columnwidth]{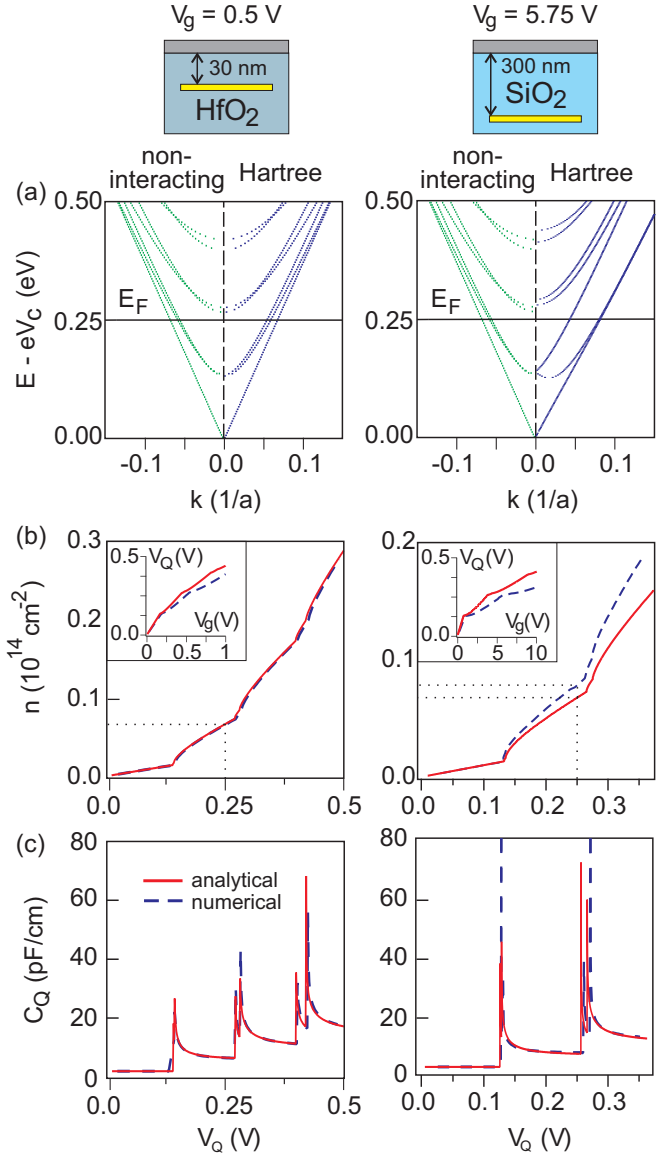}
\caption{(Color online) (a) The dispersion relations for the
non-interacting and the Hartree electrons (left and right parts of
the diagrams respectively). For the Hartree case the dispersion
relation is calculated for $V_{Q}=0.25V$; the corresponding values
of $V_{g}$ are indicated above the figure. (b) Analytical (solid
lines) and numerical (dashed lines) electron concentration $n$ and
(c) quantum capacitance $C_{Q}$ as a function of the Fermi energy
level with respect to the Dirac point ($V_{Q}$). The dotted lines
serve as a guide to compare the analytical and numerical electron
densities calculated for $V_{Q}=0.25V$. The insets in (b) show the
dependence of $V_{Q}$ on the applied gate voltage $V_{g}$. Left and
right panels corresponds respectively to HfO$_{2}$ and SiO$_{2}$
structures.} \label{fig:n(VQ)}
\end{figure}
Let us now discuss the quantum capacitance $C_{Q}$ and the
difference between the corresponding analytical and numerical
results. The crucial assumption used in the analytical model is that
an application of the gate voltage $V_{g}$ simply shifts the bands
on the amount $V_{C}$ such that the DOS of the GNR remains unchanged
relatively to the charge neutrality point for any $V_{g}$, Eq.
(\ref{mu3}). Because the amount of the induced charges is completely
determined by the value of $V_{Q}$ (Eq. \ref{n2}), this assumption
implies that the dependence $n=n(V_{Q})$ obtained by both analytical
and numerical calculations should coincide. In order to verify this
assumption we compare the evolution of the band diagram for
structures with different classical capacitances for different gate
voltages. Figure \ref{fig:n(VQ)} (a) shows the dispersion relation
for HfO$_{2}$ and SiO$_{2}$ structures where the gate voltages
$V_{g}$ are
chosen such that $V_{Q}$ is the same in both cases ($V_{Q}=0.25$V for $%
V_{g}^{\text{HfO}_{2}}=0.5$V and $V_{g}^{\text{SiO}_{2}}=5.75$V).
(For comparison we also display the dispersion relations for
non-interacting electrons). Even though the equal values of $V_{Q}$
imply the same induced charge density, the changes of the dispersion
relations are quite different. In the considered $V_{g}$ interval
the dispersion relation of the HfO$_{2}$ structure remains
practically unchanged, and therefore the analytical and the
numerical dependencies for $n=n(V_{Q})$ as well as for
$C_{Q}=C_{Q}(V_{Q})$ are almost undistinguished (see Fig.
\ref{fig:n(VQ)} (b), (c), left panel).
Modification of the band structure is much stronger for the case of the SiO$%
_{2}$ structure with smaller $C_{C}$. This is also reflected in the
analytical and numerical dependencies $n=n(V_{Q})$ exhibiting a
difference up to 15\% in the considered gate voltage interval.
However, despite of this difference for $n=n(V_{Q})$ the
corresponding difference between the analytical and numerical
results for $C_{Q}=C_{Q}(V_{Q})$ is practically negligible even for
the SiO$_{2}$ structure, see Fig. \ref{fig:n(VQ)} (b), (c), right
column.

The reason for the modification of the band structure can be
understood from the analysis of the potential distribution shown in
Fig. \ref{fig:density} (c). In order to shift $V_{Q}$ on the same
value one should apply a higher gate voltage to the structure with a
smaller classical capacitance. Different gate voltages applied to
different structures produce the same shift of $V_{Q}$ but give rise
to different distributions of the electrostatic potential across the
nanoribbon. The difference between the
electrostatic potential at the middle of the ribbon and at the edges, $%
\Delta V_{H},$ is an order of magnitude higher for the structure
with smaller $C_{C}$. Since the ribbon width $w$ is the same for
both structures, this leads to higher effective transverse electric
field $\overline{E}\sim \frac{\Delta V_{H}}{w/2}$ for the case of
SiO$_{2}$ structure, which, in turn, modifies the band structure of
the GNR. Note that the effect of the electric field on the band
structure and the DOS of graphene nanoribbons was reported before by
many authors, resulting in e.g. the energy-gap modulation
semiconductor armchair ribbons\cite{Ritter} or opening of the energy
gap for the case of zigzag ribbons\cite{Son}.

We demonstrated above that the analytical and the numerical dependencies $%
C_{Q}=C_{Q}(V_{Q})$ show an excellent agreement in the considered gate
voltage intervals both for HfO$_{2}$ and SiO$_{2}$ structures (i.e. the
utilization of Eq. (\ref{mu3}) is fully justified). However, the analytical
and numerical dependencies $V_{Q}=V_{Q}(V_{g})$ show some discrepancy, see
insets to Fig. \ref{fig:n(VQ)} (b). This, in turn, leads to a discrepancy
between the analytical and numerical $C_{Q}$ as a function of $V_{g}$ as
shown in Fig. \ref{fig:capacitance} (c). This discrepancy is manifest itself
in the difference of the gate voltage scale and not in the difference of the
magnitudes of $C_{Q}.$ Using Eqs. (\ref{V_g}) and (\ref{def_C}), the change
of $V_{Q}$ can be easily related co the change in $V_{g},$ $\frac{\partial
V_{Q}}{\partial V_{g}}=\frac{C_{tot}}{C_{Q}}=\left( 1+\frac{C_{Q}}{C_{C}}%
\right) ^{-1}.$ Because for a given $V_{Q}$ the analytical and numerical $%
C_{Q}$ are practically the same, the difference between the analytical and
numerical dependencies $\frac{\partial V_{Q}}{\partial V_{g}}$ is primarily
due to the difference of the corresponding classical capacitances $C_{C}$.
We therefore conclude that the discrepancy between the analytical and
numerical $C_{Q}$ as a function of $V_{g}$ is related to the difference in
the corresponding classical capacitances.

Let us now discuss experimental determination of the quantum capacitance $%
C_{Q}.$ In our both analytical and numerical approaches we are in position
to calculate $C_{Q}$ directly. In contrast, $C_{Q}$ is not directly
accessible in experiments. For example, for the case of carbon nanotubes it
is the total capacitance $C_{tot}$ that is measured experimentally. The
quantum capacitance $C_{Q}$ is then extracted from $C_{tot}$ according to
Eq. (\ref{def_C}), $C_{Q}^{-1}=C_{tot}^{-1}-C_{C}^{-1},$ where $C_{C}$ is a
corresponding analytical expression for the classical capacitance of a
nanotube (i.e. a classical capacitance between a metallic cylinder and an
infinite plane)\cite{Ilani,Dai}. Our calculations presented above
demonstrate that for the graphene nanoribbons the numerical $C_{C}$ differs
from its classical analytical expression given by Eq. (\ref{C_C}).
Therefore, a question arises, to what extent one can rely on the above
procedure for the extraction of the quantum capacitance?

In order to answer this question let us assume that our numerically
calculated $C_{tot}$ (shown in Fig. \ref{fig:capacitance} (b)) corresponds
to the experimental data. We then assume that the classical capacitance of
the GNR is given by the analytical expression (\ref{C_C}) describing a
capacitance between a metallic strip and an infinite plane. Finally, the
quantum capacitance (which we will call \textquotedblleft
extracted\textquotedblright , $C_{Q}^{\text{extracted}})$ is obtained by
subtracting $C_{C}$ from the \textquotedblleft measured\textquotedblright\ $%
C_{tot}$ according to Eq. (\ref{def_C}). The \textquotedblleft
extracted\textquotedblright\ quantum capacitance is shown in Fig. \ref%
{fig:capacitance} (c). For the case of the HfO$_{2}$ structure (where $%
C_{C}\lesssim C_{Q}),$ the behavior of $C_{Q}^{\text{extracted}}$ is
qualitatively similar to that of the numerical $C_{Q},$ even though their
values differ significantly. However, for the case of SiO$_{2}$ structure
(where $C_{C}\ll C_{Q}$) the \textquotedblleft extracted\textquotedblright\
quantum capacitance $C_{Q}^{\text{extracted}}$ does not reproduce the
numerical $C_{Q}$ even qualitatively with all the features related to the
quantum mechanical DOS being completely lost. This is simply related to the
fact that due to the series addition of the capacitances, for the case of $%
C_{C}\ll C_{Q}$ the total capacitance is completely dominated by the
classical one, and the features in $C_{Q}$ can be reproduced only when $%
C_{C} $ becomes comparable to $C_{Q}.$ It is interesting to note that the
difference between the experimentally extracted quantum capacitance and its
expected value was detected for the case of the carbone nanotubes and was
attributed to the strong electron correlation and signatures of the
Luttinger liquid behaviour\cite{Dai}. Our calculations indicate that the
origin of such deviations can have a rather simple explanation related to
the inability of a standard electrostatics to reproduce quantitatively the
classical capacitance of the structure at hand.

\section{Conclusions}

In the present paper we develop an analytical theory for gate electrostatics
and classical and quantum capacitance of graphene nanoribbons (GNRs). We
compare the analytical theory with the exact self-consistent numerical
calculations based on the tight-binding $p$-orbital Hamiltonian within the
Hartree approximation.

We find that the analytical theory is in a good qualitative (and in
some aspects quantitative) agreement with the exact calculations.
There are however some important discrepancies. In order to
understand the origin of these discrepancies we investigate the
self-consistent electronic structure and the charge density
distribution in the GNRs obtained from the exact numerical
calculations. We demonstrate that the assumptions appropriate for a
classical capacitor (the charge density is homogeneous and the
potential of the conductor is constant) are violated for the
graphene nanoribbons which leads to the difference between the
analytical theory and the numerical calculations. In turn, the
failure of the classical electrostatics is traced to the quantum
mechanical effects leading to the significant modification of the
self-consistent charge distribution in comparison to the
non-interacting electron pictures. We also show that as a result of
electron-electron interaction the band structure of the GNRs
modifies as the applied gate voltage increases.

Our exact numerical calculations show that the density distribution and the
potential profile in the GNRs are qualitatively different from those in
conventional split-gate quantum wires with a smooth electrostatic
confinement where the potential is rather flat and the electron density is
constant throughout the wire. At the same time, the electron distribution
and the potential profile in the GNR are very similar to those in the
cleaved-edge overgrown quantum wires (CEOQW) exhibiting triangular-shaped
quantum wells in the vicinity of the wire boundaries accompanied by the
corresponding enhancement of the electron density close to the edges. This
similarity reflects the fact that both the CEOQWs and the GNRs correspond to
the case of the hard-wall confinement at the edges of the structure.

Finally, we discuss experimental determination of the quantum capacitance $%
C_{Q}.$ We demonstrate that the extracted $C_{Q}$ might significantly
deviate from its actual value given by the density of states of the GNRs.
This deviation is related to the inability of the standard electrostatics to
reproduce quantitatively the classical capacitance of the structure at hand.

\bigskip

\acknowledgements A. A. S. and I. V. Z. acknowledge the support of the
Swedish Research Council (VR) and the Swedish Institute (SI). J. W. K.
acknowledges the support of the Polish Ministry of Science and Higher
Education within the program \textit{Support of International Mobility, 2nd
edition}.

\appendix

\section{Density of states of graphene nanoribbons}

The density of states of a quantum wire (including a factor 2 for the spin
degeneracy) reads,
\begin{equation}
\rho (E)=\left( \frac{2}{\pi }\right) \sum_{n}\left( \frac{%
dE_{n}(k_{\parallel },k_{\bot n})}{dk_{\parallel }}\right) ^{-1},
\label{Appendix_DOS}
\end{equation}%
where $k_{\parallel }(E)$ and $k_{\bot n}$ denote the longitudinal
(continuous) and the transverse (quantized) components of the wave vector,
respectively. The summation in (\ref{Appendix_DOS}) includes all transverse
modes which energy $E_{n}<E$. The dispersion relation for the nanoribbon of
the width $N$ in the low-energy limit close to the Dirac point is given by
Onipko\cite{Onipko},
\begin{equation}
E^{\sigma }(k_{\parallel })=\pm \frac{\sqrt{3}}{2}ta\sqrt{\left(
k_{\parallel }-\bar{k}_{\parallel }^{\sigma }\right) ^{2}+k_{\bot n}^{\sigma
\;2}},  \label{Appendix_E}
\end{equation}%
where $\sigma =A,Z$ corresponds to the armchair ($A$) and zigzag ($Z$) GNRs.
The transverse wave vector $k_{\bot n}^{A}$ is given by different
expressions depending on whether $\frac{2(N+1)}{3}$ is integer (metallic
nanoribbon) or not (semiconducting nanoribbon),%
\begin{equation}
k_{\bot n}^{A}=\left\{
\begin{array}{c}
\frac{\pi \left\vert n\right\vert }{N+1}\left( 1+\frac{\pi n}{4\sqrt{3}(N+1)}%
\right) a,\text{ metallic} \\
\frac{\pi }{N+1}\left( n-\frac{1}{3}\right) a,\text{ semiconducting}%
\end{array}%
\right. ,  \label{Appendix_k}
\end{equation}%
$n=0,\pm 1,\pm 2,\ldots $. For the zigzag structures, the transverse wave
vector $k_{\bot n}^{Z}$ has a form:
\begin{equation}
k_{\bot n}^{Z}=\frac{\pi (n+\tfrac{1}{2})}{2\sqrt{3}N},
\end{equation}%
$n=0,1,2,\ldots $. The parameter $\bar{k}_{\bot n}^{\sigma }$ describes the
shift of $n$-th dispersion branch minima respect to the Brillouin zone
centre,
\begin{equation}
\bar{k}_{\bot n}^{\sigma }=%
\begin{cases}
0 & \text{for $\sigma =A$} \\
\frac{2}{3}\pi +\frac{\sqrt{3}}{4}k_{\bot n}^{\sigma \;2} & \text{for $%
\sigma =Z$}%
\end{cases}%
\end{equation}%
Using Eq. (\ref{Appendix_E}) in Eq. (\ref{Appendix_DOS}), we obtain for the
DOS of the armchair (zigzag) ribbon%
\begin{equation}
\rho ^{\sigma }(E)=\frac{4}{\pi \sqrt{3}ta}\sum_{n}\frac{\left\vert
E\right\vert }{\sqrt{E^{2}-E_{n}^{\sigma \,2}}}\;\theta (\left\vert
E\right\vert -\left\vert E_{n}\right\vert ),\;
\end{equation}%
where $n=0,\pm 1,\pm 2,\ldots $ for the armchair GNRs and $n=0,1,2,\ldots $
for the zigzag GNRs, and and $E_{n}^{\sigma }=\pm \frac{\sqrt{3}}{2}%
ta\,k_{\bot n}^{\sigma }$ are the subband threshold energies. The electron
density at zero temperature is obtained by the integration of the DOS from
the charge neutrality point $\mu _{0}=0$ to the Fermi energy, $%
n=\int_{0}^{E_{F}}\rho \,dE,$%
\begin{equation}
n^{\sigma }(E_{F})=\frac{4}{\pi \sqrt{3}ta}\sum_{n}\sqrt{E_{F}^{2}-E_{n}^{%
\sigma \;2}\,}\theta (\left\vert E_{F}\right\vert -\left\vert E_{n}^{\sigma
}\right\vert ).
\end{equation}

\end{document}